\documentclass[twocolumn,showpacs,preprintnumbers]{revtex4}
\usepackage{amsfonts}
\usepackage{amsmath}

\usepackage{graphicx}
\usepackage{dcolumn}
\usepackage{bm}

\input{tcilatex}

\begin{document}

\title{Magnetic Field Induced Density of States in Superconducting MgB$_{2}$: 
Measurement of Conduction Electron Spin-Susceptibility
}
\author{F. Simon$^{1,2}$, A. J\'{a}nossy$^{1}$, T. Feh\'{e}r$^{1,2}$, F. Mur\'{a}nyi$^{1}$, 
S. Garaj$^{2}$, L. Forr\'{o}$^{2}$, C. Petrovic$^{3, *}$, S. Bud`ko$^{3}$, R. A. Ribeiro$^{3}$, P. C. Canfield$^{3}$}
\address{$^{1}$Budapest University of Technology and Economics, Institute
of Physics and Solids in Magnetic Fields Research Group of the Hungarian Academy of 
Sciences, H-1521 Budapest, PO BOX 91, Hungary\\
$^{2}$IPMC, \'Ecole Polytechnique F\'ed\'erale de Lausanne, CH-1015 Lausanne, 
Switzerland\\
$^{3}$Ames Laboratory, U.S. Department of Energy and Department of Physics\\
and Astronomy, Iowa State University, Ames, Iowa 50011}
\date{\today}

\begin{abstract}
The magnetic field dependence of the spin-susceptibility, $\chi _{s}$ was measured in the superconducting 
state of high purity MgB$_{2}$ fine powders below 1.3 T. $\chi _{s}$ was determined 
from the intensity of the conduction electron spin resonance spectra at 
3.8, 9.4, and 35 GHz. At the lowest magnetic fields (0.14 T), a gap opens in the 
density of states at the Fermi energy and, accordingly, $\chi _{s}(T)$ is small at low 
temperatures. Fields above 0.2 T (about 15 \% of $H^{c}_{c2}$, the minimum upper critical field), 
destroy the gap. The field induced $\chi _{s}$ is much larger than expected from 
current superconductor models of MgB$_{2}$.
\end{abstract}

\maketitle

It is now generally accepted that MgB$_{2}$ is a phonon mediated
superconductor\cite{akimitsu}\cite{budkoPRL}\cite{mazinPRL} which owes its
unusual properties to the widely different electron-phonon couplings on its
disconnected Fermi surface (FS) sheets\cite{mazinPRLtwogap}. The two-gap
model assumes a large superconductor gap for the cylindrical sheets
originating from B-B $\sigma $ bonds and a smaller gap for the FS sheet of $%
\pi $ electrons. It describes physical properties such as the temperature
dependent specific heat \cite{bouquetPRL} and tunneling\cite{szaboPRL}
successfully in zero magnetic field, predictions in finite fields are much
less tested. Recent experiments show that magnetic fields as low as $%
H_{c2}^{\pi }\approx $1 T close the smaller, $\pi $-gap\cite
{bouquetsingcryst}\cite{eskildsen}\cite{pointcontact}\cite{samuelyCM}. The
stronger electron-phonon coupling for $\sigma $ electron states maintains
the superconductivity above 1 T. Yet, there has been no microscopic
description of this phenomenon. The cylinder-like FS sheets of the $\sigma $
bands are responsible for the anisotropy of the upper critical field, $%
H_{c2}^{ab}=16$ $T$ and\ $H_{c2}^{c}$ about $2.5$ $T$\cite{simonPRL}\cite
{budkoPRB1}\cite{budkoPRB2}. Band calculations\cite{mazinPRLtwogap} estimate
that about half of the density of states (DOS) is on the $\pi $ band thus
the two-gap model predicts that closing the gap at 1 T restores about half
of the normal state DOS. Naively, one would expect that properties
proportional to the DOS, like\ the specific heat coefficient or the
spin-susceptibility should reach half of their normal state values at about
1 T, whereas the remaining half originating from the $\sigma $ bands should
be restored gradually with further increasing the field to $H_{c2}$.

In this Letter, we report on the magnetic field dependence of the conduction
electron spin-susceptibility, $\chi _{s}$, in the superconducting state of
high purity MgB$_{2}$ powders. This is the first time $\chi _{s}$ is
measured using conduction electron spin resonance below $T_{c}$ in a
superconductor. We observe an unusually strong increase of $\chi _{s}$ with
the magnetic field, $H$, which cannot be reconciled with current models
easily. The DOS induced by fields well below $H_{c2}$ is larger than 50 \%,
the value expected from closing the gap on the $\pi $ band FS only.

We studied samples from several batches. Sample 1 is made from 99.99 \%
purity natural isotopic mixture of $^{10}$B and $^{11}$B amorphous boron,
Samples 2 and 3 are made from crystalline, isotopically pure $^{11}$B.
Chemical analysis, the high normal state conductivities and narrow CESR
lines at $T_{c}$ attest the high purity of the samples. Details of sample
characterization are discussed in Ref. \cite{purityeffect}. Sample 2 was
made from the batch used in the CESR work of Ref. \cite{simonPRL}. Fine
powders with grain sizes less than 1 $\mu $m were selected from the starting
materials to reduce the inhomogeneity of microwave excitation. The
aggregates of small grains of the starting materials were first thoroughly
hand-crushed in a borocarbide mortar. The resulting powders were suspended
in isopropanol and mixed in a 1:1 weight ratio with pure SnO$_{2}$ fine
powder. The larger particles were eliminated by sedimentation for long times
or in a centifuge and the small grains were extracted from the suspension by
filtering. Mixing with SnO$_{2}$ is important to separate the MgB$_{2}$
particles to avoid eddy-currents. SEM microscopy has confirmed that this
procedure results in MgB$_{2}$ grains smaller than 0.5 $\mu $m size. SQUID
magnetometry of the original batch and the final fine powders confirmed that
superconducting properties were not affected by this procedure. The powders
were finally cast into epoxy for the ESR experiments. We detail CESR
experiments at 3.8, 9.4 and 35 GHz at fields near 0.14, 0.34 and 1.28 T
respectively. The CESR results at higher frequencies of Ref.\cite{simonPRL}
were also reproduced and we confirmed that the minimum value of the higher
upper critical field, $H_{c2}^{min}$, is somewhat below 2.7 T. $\chi _{s}$
of MgB$_{2}$ was measured from the intensity of the CESR with no need for
core electron corrections. Fine powder of the air stable metallic polymer, $%
o $-KC$_{60}$ was mixed into the epoxy for some of the samples to serve as a
temperature independent ESR intensity standard \cite{bommeliPRB}. Intensity
measurements at 9 GHz (where instrumental factors are well controlled) with
and without KC$_{60}$ were consistent. The absolute value of $\chi _{s}$ was
measured against a secondary CuSO$_{4}\cdot $5 H$_{2}$O standard at 9.4 GHz.

\begin{figure}[tbp]
\includegraphics[width=0.6\hsize]{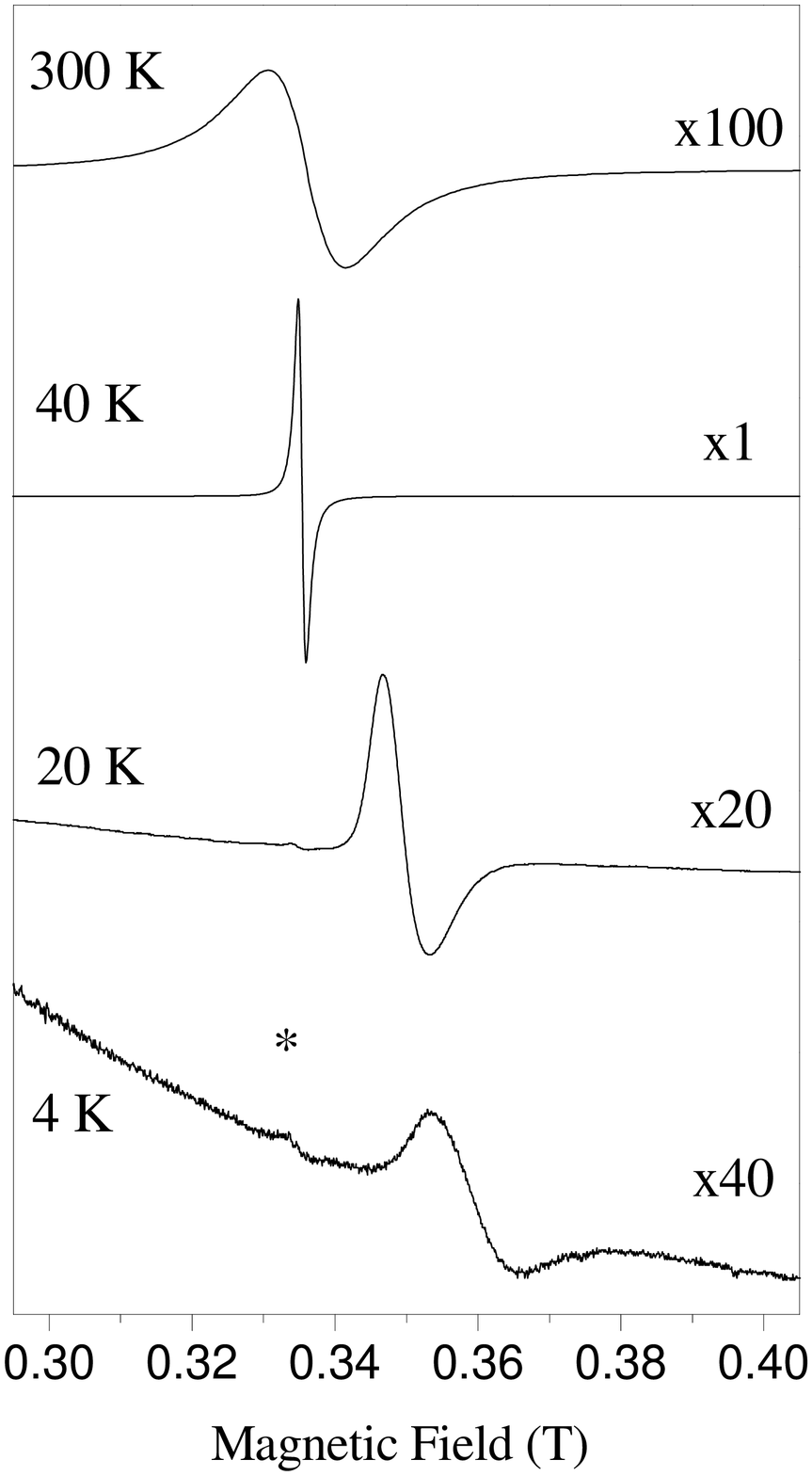}
\caption{ Temperature dependence of the CESR signal of MgB$_{2}$ fine powder
sample (Sample 1) at 9.4 GHz (0.34 T). * denotes the ESR signal of a tiny
amount of paramagnetic impurity phase}
\label{spectra}
\end{figure}

Figure 1 shows the CESR spectra at 9.4 GHz for Sample 1. The room
temperature peak-to-peak linewidth, $\Delta H_{pp}$=111$\pm 3$ G, of the
derivative absorption line is dominated by spin-lattice relaxation due to
phonons. The residual linewidth at 40 K arises from static imperfections
and is sample dependent. The residual linewidth is somewhat smaller for
Sample 1 ($\Delta H_{pp}$=10$\pm 0.3$ G), than for Sample (2,3), (11$\pm 0.3$
and 20$\pm 0.6$ G) while the residual resistance ratio (RRR)\ is larger for
the latter two samples \cite{purityeffect}. Unlike the RRR, the CESR
linewidth is insensitive to intergrain scattering and the difference may be
related to the different morphology of Sample 1 and (2,3). The small
asymmetry of the lineshape ($A/B$=1.16 for Sample 1) at 40 K shows that
microwave penetration is nearly homogeneous\cite{dyson}, the nominal
particle size is in the range of the microwave penetration depth, $\delta $%
=0.3 $\mu $m, and the reduction of the CESR\ signal intensity due to
screening is less than 5 \%.

Above 450 K (data not shown), the CESR intensities are the same for the
three samples and correspond to a susceptibility of $\chi _{s}$=(2.3$\pm $%
0.3)$\cdot $10$^{-5}$ emu/mol in agreement with the previously measured
value\cite{simonPRL} of $\chi _{s}$=(2.0$\pm $0.3)$\cdot $10$^{-5}$ emu/mol and
calculations of the DOS\cite{mazinPRL}. In Sample 1, the CESR intensity is
nearly $T$ independent in the normal state, the measured small decrease of
about 20 \% between 600 and 40 K is of the order of experimental precision at high temperatures.
In Sample 2 and 3 the CESR intensity decreased by a factor of 2.5 between
450 and 40 K. The nearly symmetric Lorentzian lineshape showed that this
intensity decrease is not due to a limited penetration depth. Neither does
the intensity decrease correspond to a change in $\chi _{s}$. We measured 
the $T$ dependence of the $^{11}$B spin-lattice relaxation
time, $T_{1}$, in Samples 2 and 3 and found a metallic, $T$ independent
value of $1/(TT_{1})$= 167$\pm 3$ s$^{-1}$ in agreement with Ref. \cite
{borsaPRB} and \cite{kotegawa}. It is possible that the difference in
morphology and purity at the grain surfaces explains that the CESR signal
intensity is almost constant in Sample 1 while it changes strongly in Sample
2 and 3. We believe that the nearly constant CESR intensity and the constant 
$1/(TT_{1})$ measures correctly the metallic susceptibility of MgB$_{2}$.

\begin{figure}[tbp]
\includegraphics[width=1\hsize]{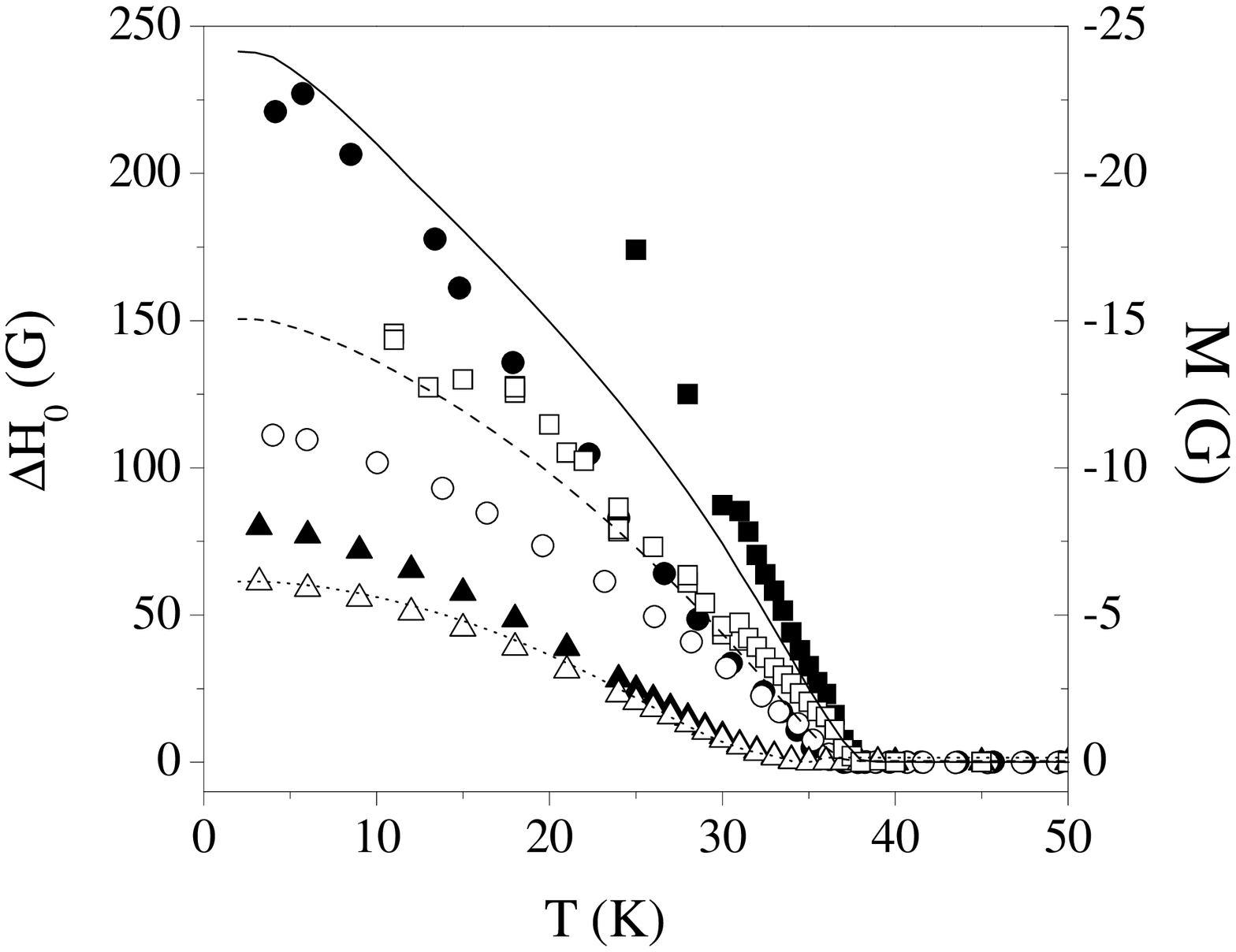}
\caption{ Temperature dependence of the diamagnetic shift (full symbols are
up, open symbols are down sweeps) of the CESR (squares: 3.8 GHz, 0.14 T; circles: 9.4 GHz
0.34 T; triangles: 35 GHz, 1.28 T) and diamagnetic magnetization
measured by SQUID (solid, dashed, and doted curves are at 0.14, 0.34, and 1.28 T, respectively.}
\label{diam}
\end{figure}

Below $T_{c}$, the $T$ and $H$ dependence of the diamagnetic shifts, linewidths, 
and intensities normalized at $T_{c}$ are similar in all three
samples. We discuss Sample 1, for which the CESR signal intensity is
constant in the normal state within experimental precision. For the applied
magnetic fields, $H<H_{c2}^{min}(T=0$ K$)$ and at $T\ll T_{c}$ the CESR
signal corresponds to the mixed state of the MgB$_{2}$ superconductor; any
non-superconducting fraction would be easily detected as in Ref.\cite
{simonPRL} for $H>H_{c2}^{min}$. The ESR of a tiny impurity phase is well
outside the CESR of MgB$_{2}$ (marked by * in Fig. 1). Comparison of the
diamagnetic magnetization, $M$, measured by SQUID and the diamagnetic shift
of the CESR, $\Delta H_{0}(T)=H_{0}(T)-H_{0}(40\,${\rm K}$)$ ($H_{0}$ is the
resonance field), also verifies that below $T_{c}$ we detect the CESR of the
MgB$_{2}$ superconductor. Figure 2 shows $\Delta H_{0}(T)$ at 3 different
ESR frequencies and $M$ at the corresponding static magnetic fields with a
scaling between $\Delta H_{0}$ and $M$ as in Ref.\cite{simonPRL}. The
present diamagnetic shift data on fine grain samples at 35 GHz and higher
frequencies (not shown) agrees with our previous report on large grain
samples\cite{simonPRL}. $\Delta H_{0}(T)$\ is equal to the average decrease
of the applied magnetic field in the sample and is proportional within a
shape dependent constant to $M$ of the grains. $\Delta H_{0}(T)$ averaged
for increasing and decreasing field sweeps is proportional to $M$ in the
field cooled sample with the same proportionality constant in a broad range
of $T$'s for all the three magnetic fields. We used the same argument
previously to identify the signal of superconducting MgB$_{2}$ particles in
higher frequency (>35 GHz) CESR experiment\cite{simonPRL}.

\begin{figure}[tbp]
\includegraphics[width=1\hsize]{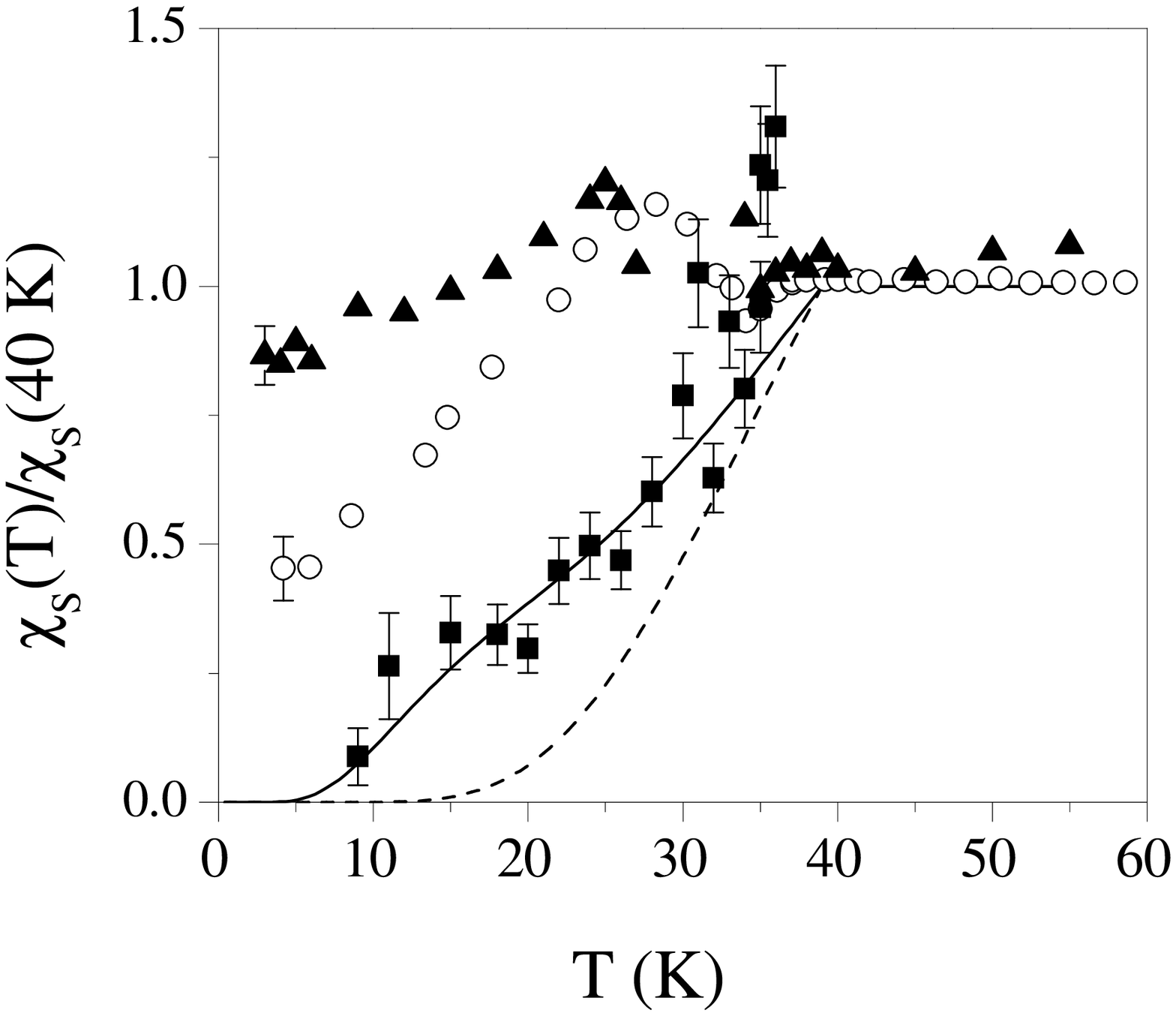}
\caption{The temperature and magnetic field dependent spin-susceptibility of
MgB$_{2}$ below $T_{c}$ (squares: 3.8 GHz, 0.14 T;circles: 9.4 GHz 0.34 T;
triangles: 35 GHz, 1.28 T). Dashed and solid curves are calculated $\protect%
\chi_{s}$ for an isotropic BCS and a two-gap model as explained in the text,
respectively.}
\label{susc}
\end{figure}

The CESR line broadens inhomogeneously below $T_{c}$ due to the macroscopic
inhomogeneities of diamagnetic stray fields (data not shown). Within each
isolated grain, spin diffusion motionally averages all magnetic field
inhomogeneities arising from screening currents or vortices\cite{degennes}
and the modulation of field between vortex cores does not broaden the CESR.
The line is however broadened in the randomly oriented powder by the
crystalline and shape anisotropy of the diamagnetic shift, that changes from
grain to grain. As expected for this case, the observed additional linewidth
in the superconducting state, $\Delta H_{a}$, is proportional to the shift
and $\Delta H_{a}/\Delta H_{0}(T)$ is about 0.3 at all fields and
temperatures.

The main topic of the current report is the measurement of $\chi _{s}$ in
the superconducting state of MgB$_{2}$. Usually, $\chi _{s}$ is determined
from the measurements of temperature dependent Knigths shift or spin-lattice
relaxation time, $T_{1}$\cite{slichter}. To our knowledge, in MgB$_{2}$
there has been no successful determination of these quantities below $T_{c}$%
: precision of the Knigth shift measurement is limited due to the
diamagnetic magnetization\cite{borsaPRB}\cite{gerashenko}, while $T_{1}$
measurements are affected by several factors like vortex motion\cite
{borsaPRB}. In CESR, the signal intensity is proportional to $\chi _{s}$ and
diamagnetism affects only the shift of the resonance line. $\chi _{s}$ is
proportional to the DOS of normal excitations when electron correlations are
small.

In Figure 3, we show $\chi _{s}$ below 60 K measured at three magnetic
fields. The 9.4 GHz (0.34 T) and 35 GHz (1.28 T) data are normalized at 40 K
while the 3.8 GHz (0.14 T) data are normalized at 100 K. At 3.8 GHz and 35
GHz data are missing in ranges of 36 to 100 K and 27 to 34 K, respectively,
where the CESR of MgB$_{2}$ could not be resolved from the KC$_{60}$
reference. Data taken at 9.4 GHz with and without reference agree well and
the data without reference are shown. Irreversibility does not affect the
measurement of $\chi _{s}$ in the studied range of $T$ and $H$: linewidths
and shifts depend on the direction of the field sweep at low $T$ but the
intensities are the same within 5 \% that is our experimental precision at low $T$. In
Fig. 3, data at 9.4 and 35 GHz are averaged for sweeps with increasing and
decreasing fields while at 3.8 GHz decreasing field sweep data are shown. No
correction is made for diamagnetic screening of the microwave excitation,
this would increase somewhat further the measured values of $\chi _{s}$.

The $T$ dependent $\chi _{s}$ at our lowest magnetic field, 0.14 T, lies
well above $\chi _{s}^{0}(T,H=0)$ (dashed curve in Fig. 3) calculated for an
isotropic $T_{c}$ =39 K weak coupling BCS\ superconductor\cite{yosida}. The
measured $\chi _{s}$ at 0.14 T is compatible with the two gap model above 10
K. The solid curve in Fig. 3 shows $\chi _{s}(T,H=0)$ calculated within the
model of two independent gaps. We used $\Delta _{1}(T=0)$=11 K\ and $\Delta
_{2}(T=0)$=45 K, with DOS equally shared on the two Fermi surfaces, and both
gaps opening at $T_{c}$. The choice of these parameters is in agreement with
experimental results \cite{bouquetPRL}\cite{szaboPRL}\cite{junodbouquetEPL}
and theoretical estimates\cite{mazinPRLtwogap}.

\begin{figure}[tbp]
\includegraphics[width=1\hsize]{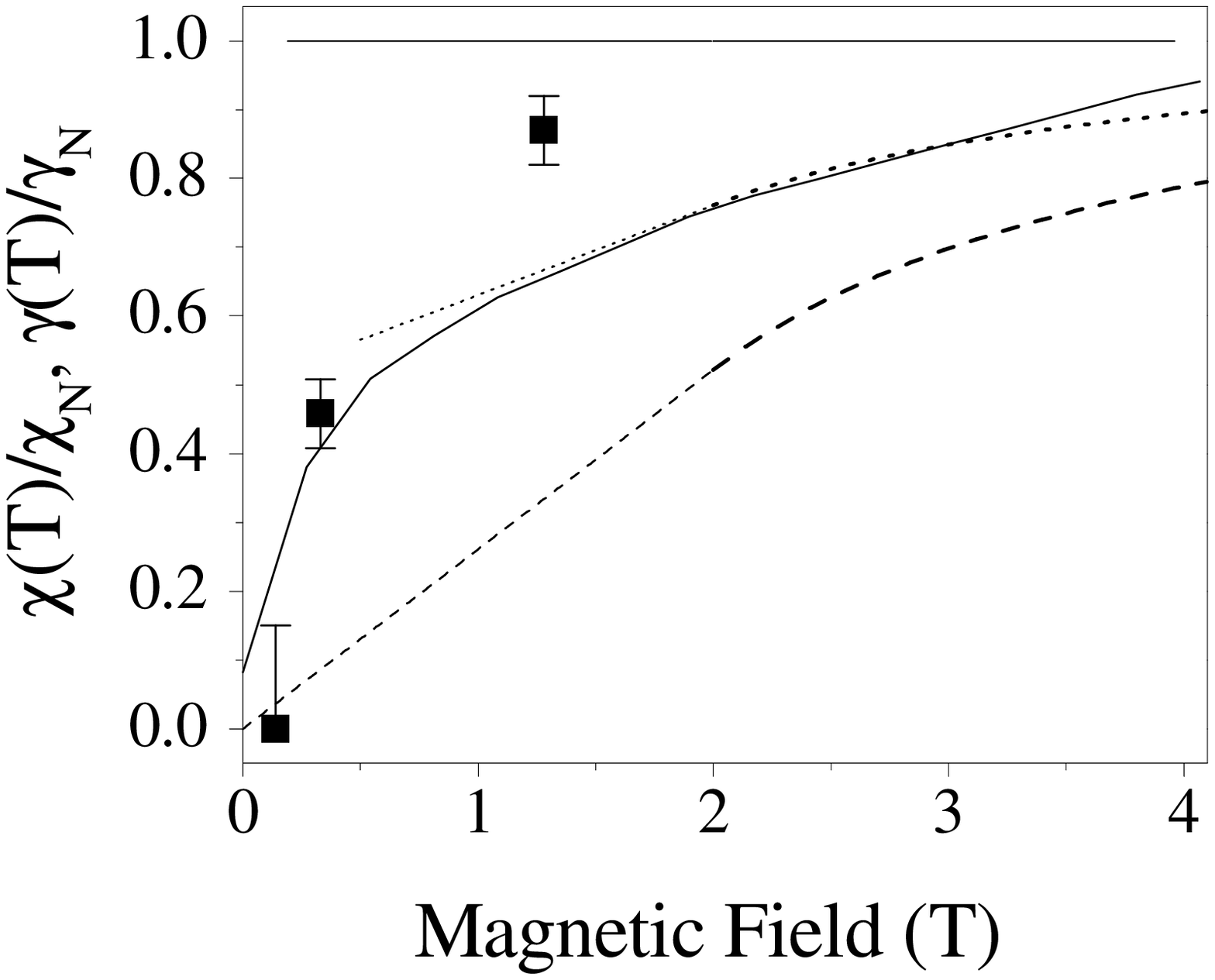}
\caption{Comparison of the normalized $\protect\chi_{s}/\protect\chi_{N}$ at
4 and 3 K at 0.34 and 1.28, respectively (squares) and the normalized
specific heat coefficient $\protect\gamma/\protect\gamma_{N}$ (solid line)
at 3 K from Ref. \protect\cite{junodbouquetsumm} . Dashed curve shows $H$
dependent DOS for an anisotropic $H_{c2}^{c}=2$ T, $H_{c2}^{ab}=16$ T
superconductor. Dotted curve is a similar superconductor, but with half of
the DOS restored by 0.5 T. For 16 T, all curves merge at 1.}
\label{spec_heat}
\end{figure}

The remarkably large field induced $\chi _{s}$ observed at 0.34 and 1.28 T
at all $T$'s below $T_{c}$ is the main finding of this work. These fields
are much smaller than the upper critical fields, $H_{c2}^{c}\sim $2.5 T \cite
{simonPRL}\cite{budkoPRB1}\cite{budkoPRB2} and $H_{c2}^{ab}$ $\sim $ 16 T.
Unlike in usual superconductors, $\chi _{s}$ at $T_{c}/2$ is still near to its
normal state value and remains large at the lowest $T$'s; $\chi _{s}$(4 K)/$%
\chi _{s}$(40 K) = 0.46$\pm $0.06 at 0.34 T and $\chi _{s}$( 3 K)/$\chi _{s}$%
( 40 K) =0.87$\pm $0.06 at 1.28 T. As shown in Fig. 4, the measured $\chi
_{s}$ at low $T$ is in rough agreement with the specific heat\cite
{bouquetPRL}\cite{junodbouquetsumm} measured on a powder sample (solid curve
in Fig. 4) but we find an even stronger field dependence. The specific heat data, unlike
the $\chi _{s}$ data, are affected by the electron-phonon enhancement which varies
strongly between the different FS sheets. The data cannot be
described by a simple anisotropic superconductor. The dashed curve in Fig. 3
shows the calculated field dependence of $\chi _{s}$ of a powder of an
anisotropic superconductor with $H_{c2}^{c}$=2 T and $H_{c}^{ab}$=16 T at $T$%
=0. Here, we assumed a Landau-Ginzburg type angular dependence of $%
H_{c2}(\theta )$ as described in Ref.\cite{simonPRL} and a contribution to $\chi _{s}$ proportional to $H/H_{c2}$%
($\theta $) for each superconducting grain. As seen in Fig. 4, this
description is inadequate. A better approximation is obtained if, following
Ref. \cite{bouquetsingcryst} \cite{eskildsen}\cite{pointcontact}, one
assumes that the $\pi $ band FS is restored at around $H\sim $1 T and the
observed anisotropic $H_{c2}$ arises solely from the $\sigma $ band. In Fig.
4 we show this case with dotted lines, assuming that 50 \% of the DOS is
already restored at $H=0.5$ T. However, this simple minded application of
the two-gap model is still inadequate: the large value of $\chi _{s}$ at
1.28 T and 3 K, and the $T$ dependence at 0.34 T, in particular the
restoration of the full DOS at 25 K, remain unexplained.

In conclusion, the magnetic field dependence of the spin-susceptibility in
MgB$_{2}$ shows that a large part of the Fermi surface is restored at fields
well below the minimum of the upper critical field. These observations pose
a challange to theory since they cannot be explained with an anisotropic
superconductor model nor by simple application of the two-gap model even if
it is assumed that the $\pi $ band FS is restored at magnetic fields as low
as 0.1$\cdot H_{c2}^{\min }$.

Support from the Hungarian State Grants, OTKA T029150, OTKA TS040878, and
FKFP 0352/1997 and the NCCR network MANEP of the SNSF are acknowledged. The
authors are grateful to Silvija Grade\v cak and Csilla Mik\'o for the SEM
experiments and to Edina Couteau for assistance in the fine powder preparation. 
FS and TF acknowledge the Bolyai Hungarian Research Fellowship
for support. TF acknowledges the hospitality of the Grenoble High Magnetic Field 
Laboratory during the NMR measurements. Ames Laboratory is operated for the U.S. 
Department of Energy by Iowa State University under Contract No. W-7405-Eng-82.

$^{*}$ Present address: Department of Physics, Brookhaven
National Laboratory, Upton, New York 11973

\end{document}